\documentclass[reprint,aps,pre,amsmath,10pt,twocolumn,superscriptaddress,showpacs]{revtex4-1} 

\usepackage{graphicx}
\usepackage{subfigure}
\AtBeginDocument{}    

\begin{document}

\title{Synchronization and long-time memory in neural networks with inhibitory hubs and synaptic plasticity}
\author{\small Elena Bertolotti}
\email[Electronic address: ]{elena.bertolotti1@fis.unipr.it}
\affiliation{\footnotesize \textit{Dipartimento di Fisica e Scienza della Terra, Universit\`a di Parma, via G.P.~Usberti, 7/A-43124, Parma, Italy}}
\affiliation{\footnotesize \textit{INFN, Gruppo Collegato di Parma, via G.P.~Usberti, 7/A-43124, Parma, Italy}}
\author{\small Raffaella Burioni}
\email[Electronic address: ]{raffaella.burioni@fis.unipr.it}
\affiliation{\footnotesize \textit{Dipartimento di Fisica e Scienza della Terra, Universit\`a di Parma, via G.P.~Usberti, 7/A-43124, Parma, Italy}}
\affiliation{\footnotesize \textit{INFN, Gruppo Collegato di Parma, via G.P.~Usberti, 7/A-43124, Parma, Italy}}
\author{\small Matteo di Volo}
\email[Electronic address: ]{mdivolo@iupui.edu}
\affiliation{\footnotesize \textit{Group for Neural Theory, Depart\'{e}ment des Etudes Cognitives, Ecole Normale Sup\'{e}rieure, Paris, France}}
\affiliation{\footnotesize \textit{Centro Interdipartimentale per lo Studio delle Dinamiche Complesse, via Sansone, 1-50019 Sesto Fiorentino, Italy}}
\affiliation{\footnotesize \textit{Indiana University-Purdue University, 420 University Blvd., Indianapolis, Indiana 46202, USA}} 
\author{\small Alessandro Vezzani}
\email[Electronic address: ]{alessandro.vezzani@fis.unipr.it}
\affiliation{\footnotesize \textit{Dipartimento di Fisica e Scienza della Terra, Universit\`a di Parma, via G.P.~Usberti, 7/A-43124, Parma, Italy}}
\affiliation{\footnotesize \textit{IMEM--CNR, Parco Area delle Scienze 37/A-43124 Parma, Italy}}

\begin{abstract}
\footnotesize 
  We investigate the dynamical role of inhibitory and highly connected nodes (hub) in synchronization and input processing of leaky-integrate-and-fire neural networks with short term synaptic plasticity. We take advantage of a heterogeneous mean-field approximation to encode the role of network structure and we tune the fraction of inhibitory neurons $f_I$ and their connectivity level to investigate the cooperation between hub features and inhibition. We show that, depending on $f_I$, highly connected inhibitory nodes strongly drive the synchronization properties of the overall network through dynamical transitions from synchronous to asynchronous regimes. Furthermore, a metastable regime with long memory of external inputs emerges for a specific fraction of hub inhibitory neurons, underlining the role of inhibition and connectivity also for input processing in neural networks.
\end{abstract}

\maketitle

{\small
\section{\small INTRODUCTION}
Neural ensembles feature a wide range of dynamical patterns, that can be suitably analyzed through collective observables, like local field potentials generated by single units firing activity  \cite{eeg, fmri, oscill}. These observables typically display oscillations on different frequency ranges, that encode specific structures of synchronization in the activity of single neurons \cite{syncheeg1, syncheeg2}. Synchronization among neurons has been shown to be relevant for normal neural functions and also to be associated, when anomalous, with neural disorders \cite{synchpat1, synchpat2}. Therefore neural synchronization has been so far widely investigated, both from experimental and theoretical points of view \cite{pik, synchexp}.

Several features of neural networks have been shown to influence synchronization of neural populations, e.g.,~neural excitability \cite{synchexc}, the presence of inhibitory component \cite{synchin1} and the structure of connections \cite{synchstr1}.
In recent years, a crucial role of inhibitory component in neural ensembles has been proposed to reproduce specific patterns observed in cortical regions of the brain. Indeed, an optimal balance between excitatory and inhibitory neurons is required to obtain patterns of activity observed in different experimental setups, and an optimal fraction of inhibitory components has also been suggested to optimize the performances of a neural network \cite{inhvan, inhtask}. 

Also the structure of connections plays a central role in neurons synchronization and in their capacity to store and process information \cite{strinput}. Recently, evidence of hubs neurons in the brain, i.e.,~neurons with high connectivity, has been put forward \cite{hub1}. These neurons are crucial in organizing the synchronization of neurons they are connected with. Moreover, the activity of a single hub neuron can strongly influence the firing pattern of the overall population. A recent paper by Bonifazi \textit{et al}.~\cite{bonifazi2009_inhibhubs} has put into evidence that hub neurons are typically inhibitory, suggesting an unifying view of cooperation between inhibition and connectivity structure as a driving of synchronization properties in neural networks. 

In this paper we investigate theoretically and numerically the role and the interplay of the fraction of inhibitory neurons and of their hub character (their relative connectivity with respect of the rest of the network units) in synchronization and functional properties of a neural network. In particular, we consider here a leaky-integrate-and-fire (LIF) neural network composed by inhibitory and excitatory neurons with a short term synaptic plasticity mechanism, developed by Tsodyks and colleagues (Refs.~\cite{tsodyks1998_TUM2, tsodyks2000_TUMdef, tsodyks1997_TUM1}).
 In order to investigate the interplay between inhibition and connectivity, we tune two main parameters: the fraction of inhibitory neurons in the network, $f_I$, and their hub character quantified by the distance between the average connectivity $\langle k \rangle$ of inhibitory ($I$) and excitatory ($E$) populations, $\Delta= \langle k_I \rangle - \langle k_E \rangle$.
We show how the interplay of these two ingredients gives rise to a wide range of dynamical regimes, with specific synchronization properties and different ability to process external inputs. The network dynamics is typically organized in a quasi--synchronous regime for very small fractions of inhibitory neurons and turns to a completely asynchronous state when the network is mainly inhibitory. Nevertheless, increasing the fraction of inhibitory components does not simply decrease neural synchronization. Indeed, the synchronization of the network increases for intermediate values of $f_I$, reaching an almost synchronous regime for an optimal balance between excitation and inhibition. Furthermore, close to this balance, we find an intriguing metastable dynamical regime, which turns out to be the most effective for storing information of external inputs. 
As a result of this analysis, the fraction of inhibitory neurons and their hub character is shown to combine non trivially for the emerging network synchronization and input processing capability.

The model we consider is described in details in Sec.~\ref{sec2}, where we also implement a heterogeneous mean-field model (HMF) reproducing the finite size dynamics for sufficiently high connectivity (see, e.g., \cite{vespignani2008_libro, dorogovtsev2008_HMF}). Such a HMF model turns out to be very useful to represent the role of connectivity distribution and to reduce the computational cost of numerical simulations. An extended analysis of the HMF model ability to reproduce finite size dynamics is then reported in Sec.~\ref{sec6}. In Sec.~\ref{sec3} we discuss the synchronization properties of the network as a function of $f_I$ and of the hub character of inhibitory neurons $\Delta$ and we analyze the dynamical transitions from partially synchronous, synchronous, and asynchronous regimes. In Sec.~\ref{sec4} we describe the dynamical features of the metastable regime emerging for specific values of $f_I$ and $\Delta$. In Sec.~\ref{sec5} we investigate how the network processes external inputs, showing a high capacity of storing information in the metastable regime. Section \ref{sec6} is devoted to the comparison between the HMF results and finite size networks simulations, to show that the two dynamics become similar by increasing network connectivity, as expected. However, this limit is not trivial, since in the quasi-synchronous and asynchronous regimes the effect of a finite connectivity is a noise superimposed to the HMF behavior, while in the balance regime the HMF dynamics is obtained considering metastable states in finite size networks, whose lifetime diverges with the system connectivity. Eventually, in Sec.~\ref{sec7}, we discuss conclusions and perspectives.

\section{\small FROM THE FINITE SIZE MODEL TO THE HMF FORMULATION}
\label{sec2}
We consider a neural network where each node represents an excitatory or inhibitory neuron and directed links model the synapses connecting the corresponding nervous cells. 
Neural dynamics is described by a LIF oscillator (Refs.~\cite{gerstner2002_libro_LIF, izhikevich2007_libro}), so that the evolution of the membrane potential $v_i(t)$ of neuron $i$ is given by
\begin{equation}
\label{eqv}
\dot{v}_{i}\,=\,a - v_i + I^{syn}_{i}(t),
\end{equation}
where $a$ is the leakage current and $ I^{syn}_{i}(t)$ is the synaptic current, encoding the coupling with other neurons. Without loss of generality we have rescaled variables and parameters, that are now expressed in adimensional units. Whenever the membrane potential $v_{i}$ reaches the threshold $v_{i}^{th}\,=\,1$, a spike is instantaneously sent to all the postsynaptic connections of neuron $i$ and $v_i$ is reset to $v_{i}^{r}\,=\,0$. For $a > 1$ the neuron emits periodically spikes even if it does not interact with the rest of the network and it is said to evolve in a spiking regime.

We describe the coupling among neurons with the Tsodyks-Uziel-Markram (TUM) model for short term synaptic plasticity, already presented in \cite{tsodyks2000_TUMdef}. Each synapse connecting neuron $j$ to neuron $i$ is endowed with synaptic transmitters or resources, that can be found in three different states. Accordingly, for each synapse $(i,j)$ a fraction of available, active, and inactive resources, $x_{ij}$, $y_{ij}$, $z_{ij}$, respectively, is defined and their time evolution equations read
\begin{gather}
\label{eqy}
\dot{y}_{ij}(t)\,=\, - \frac{y_{ij}(t)}{\tau_{in}}\,+\,u_{ij}(t) x_{ij}(t) S_{j}(t),\\
\label{eqx}
\dot{x}_{ij}(t)\,=\, \frac{z_{ij}(t)}{\tau_{r}^{i}}\,-\,u_{ij}(t) x_{ij}(t) S_{j}(t),\\
\label{norm}
x_{ij}(t)\,+\,y_{ij}(t)\,+\,z_{ij}(t)\,= \,1,
\end{gather}
where the last equation is the resources normalization and $S_{j}(t)$ represents the spike train produced by the presynaptic neuron $j$: 
\begin{equation}
S_{j}(t)\,=\,\sum_{n} \delta [t - t_{j}(n)],
\end{equation}
where $t_{j}(n)$ is the time at which neuron $j$ emits its \textit{n}th pulse. In between two consecutive spikes by neuron $j$ the active resources $y_{ij}$ exponentially decay with a time constant $\tau_{in}$, while the inactive resources $z_{ij}$ are recovered, returning available in a time $\tau_{r}^{i}$. Actually, when neuron $j$ spikes, it activates a fraction $u_{ij}$ of the available resources $x_{ij}$. If the postsynaptic neuron $i$ is excitatory, $u_{ij}=U$ is constant in time. Otherwise, if $i$ is inhibitory, $u_{ij}$ evolves according the following facilitation mechanism:
\begin{equation}
\label{equ}
\dot{u}_{ij}(t) = - \frac{u_{ij}(t)}{\tau_{f}}\,+\,U_f\,[1 - u_{ij}(t)]S_{j}(t).
\end{equation}
Accordingly, the variable $u_{ij}$ decays in a time $\tau_{f}$, while it increases instantaneously of an amount $U_f$ at the spiking event. Moreover, for postsynaptic inhibitory neurons, $\tau_{r}^{i}$ is typically much smaller than in the excitatory case.
The equations are then closed by the synaptic current $I^{syn}_i(t)$ in Eq.~(\ref{eqv}), defined as the sum over all the connected active resources $y_{ij}$: 
\begin{equation}
\label{sincurr}
I^{syn}_{i}(t)\,=\,\frac{g}{\langle k \rangle}\,\sum_{j \neq i} \epsilon_{ij} y_{ij}(t),
\end{equation}
where $g$ is the strength of the synaptic coupling and $\epsilon_{ij}$ is the connectivity matrix ($\epsilon_{ij} = 0$ if there are no links between $i$ and $j$, $\epsilon_{ij} = 1$ if the presynaptic neuron $j$ is excitatory, $\epsilon_{ij} = -1$ if $j$ is inhibitory). 
The number of synapses outgoing from neuron $j$ is $k_{j,out} = \sum_i |\epsilon_{ij}|$. Similarly, the input degree $k_{i,in} = \sum_j |\epsilon_{ij}|$ corresponds to the number of neurons from which $i$ receives synaptic inputs. Accordingly, $\langle k \rangle$ is the average in-degree of the graph. Equations (\ref{eqv})-(\ref{sincurr}) can be very efficiently simulated through an event-driven map, exploiting the integrability of the model between two subsequent firing events (see, e.g., Refs.~\cite{brette2006_eventdriven, livi2007_primo}).

In this paper we will study the case where the inhibitory neurons are hubs of the networks, i.e., their connectivity is much larger than the average. 
We are interested in networks with hubs displaying a large number of both in and out connections. Accordingly, we impose a simple correlation between in- and out-degree, so that $k_{i,out}\, =\,k_{i, in}=k_i$, for each node $i$. In particular we consider a ``configuration model,'' i.e., random networks with fixed degree distribution. As we are interested in networks with very large nodes degree, that is with $k_i \rightarrow \infty$, we can approximate the discrete degree $k_i$ with a continuous variable $k$ and we can call $P_{E}(k)$ [$P_{I}(k)$] the degree p.d.f.~of excitatory (inhibitory) neurons. As a consequence, the fraction of excitatory (inhibitory) neurons $f_E$ ($f_I$) is
\begin{equation}
\label{f_I}
\begin{split}
f_E&= \int P_{E}(k) dk, \\
f_I&= \int P_{I}(k) dk,
\end{split}
\end{equation}
 where $f_E+f_I=1$ and the connectivity p.d.f.~for a generic neuron is $P_E(k)+P_I(k)$. 

For purely excitatory networks numerical simulations show, as a function of $g$ and $\tau_{in}$, different dynamical phases ranging from an asynchronous regime to a partially and a totally synchronous one \cite{parma2013_primo}. In particular, in the partially synchronous state, all neurons in the network arrange into two subgroups: the locked ones, with a periodic dynamics and a common period, and the unlocked ones, with an aperiodic dynamics and different periods. The only topological property, apart from finite size fluctuations, that surprisingly distinguishes a locked neuron from an unlocked one is its input degree $k_{i, in}$.
  
Figure \ref{fig:primo raster&isim} shows that the same property is preserved for networks with both excitatory and inhibitory neurons, as also shown in \cite{parma2016_HMF3inhib}.
Data are obtained from a network of $N=5000$ neurons, $f_I=0.1$, where both $P_E(k)$ and $P_I(k)$ are Gaussian with standard deviation $\sigma = 10$ and average values $\langle k_E \rangle = 100$ and $\langle k_I \rangle = 350$ respectively.
In the inset, the raster plot represents the spike event of each node as a function of time and nodes are ordered according to their degree $k_i$. We observe that the excitatory neurons (blue data in Fig.~\ref{fig:primo raster&isim}) with small connectivity $k_i$ are actually characterized by a periodic and phase-locked dynamics, while all the other nodes evolve aperiodically (excitatory nodes with $k_i > 115$ and all inhibitory nodes). 
The dynamical properties of the \textit{i}th neuron are synthetically represented by the time average $\langle$ISI$\rangle$ of the interspike interval (ISI is defined as the time lapse between two consecutive firing events). Plotting $\langle$ISI$\rangle$ as a function of the node degree, locked or periodic excitatory neurons with low degree (i.e., with $k_i \lesssim \langle k_E \rangle$ ) belong to the initial horizontal plateau while, for all other aperiodic neurons, $\langle$ISI$\rangle$ depends on $k_i$. Therefore, also in this framework, the dynamical properties of a neuron seem to depend on the network topology, apart from finite size fluctuations, only through the degree of the node.

\begin{figure} 
\centering
\includegraphics[width=1\columnwidth]{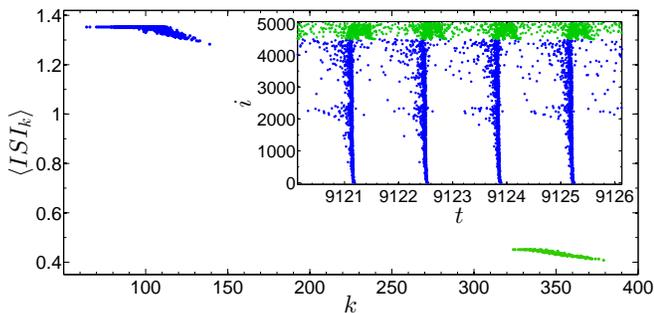}
\caption{\footnotesize{$\langle$ISI$\rangle_k$ as a function of the degree $k$ for a network of $N = 5000$ neurons, which evolves according Eqs.~(\ref{eqv})-(\ref{sincurr}). In the inset we show a raster plot of the corresponding dynamics, where nodes are ordered along the vertical axis according to their degree $k$; a dot means that neuron $i$ has fired at time $t$. Here and in all other plots time is dimensionless. The inhibitory fraction $f_I$ is $0.1$. Blue (green) data refer to the excitatory (inhibitory) nodes. The excitatory average degree $\langle k_E \rangle$ is $100$, the inhibitory one $\langle k_I \rangle$ is $350$ and the standard deviation $\sigma$ of both Gaussian p.d.f.s $P_E(k)$ and $P_I(k)$ is $10$.}}
\label{fig:primo raster&isim}
\end{figure}

We thus expect that a heterogeneous mean-field approach (Refs.~\cite{parma2014_HMF1, parma2014_HMF2}), which actually gets rid of connectivity patterns and takes into account only the neurons connectivity $k_i$ and the in-degree distribution, could be effectively applied also in our setup. 
The field $Y_{i}(t) $ received by neuron $i$ can be written as
\begin{equation}
\label{yrewrite}
    Y_{i}(t) = \frac {1} {\langle k \rangle}\sum_{j=1}^{N} \epsilon_{ij} y_{ij}(t) = \frac {k_{i}} {\langle k \rangle} \frac {1} {k_{i}} \sum_{\substack{j = 1\\ \langle i,j \rangle}}^{k_i} { y_{j}(t)},\\
 \end{equation}
 where we have put into evidence the degree of the \textit{i}th neuron. The notation $\langle i,j \rangle$ below the second sum means that only the couples where $i$ and $j$ are nearest-neighbor are now included. Indeed, each row of the matrix $\epsilon_{ij}$ is a vector with $(N-k_i)$ elements equal to $0$ and $k_i$ elements equal to $1$ and here we want to consider only the $k_i$ nonzero elements, that is those $j$ indices such that $\epsilon_{ij} = 1$.
Then we make a mean-field approximation on the right-hand side of Eq.~(\ref{yrewrite}), i.e., we approximate the average over the $k_i$ neighboring sites of neuron $i$ with the average over the whole graph. We then take into account that the contribution of each neuron has to be weighted with its out-degree, that is, in a graph with $k_{i,out}=k_{i,in}=k_i$, we need to multiply each microscopic field $y_i$ by $k_i/\langle k \rangle$.
Accordingly, we can write
 \begin{equation}
    Y_{i}(t) \approx \frac {k_{i}} {\langle k \rangle} \frac{1} {N} \sum_{j = 1}^{N} \frac {k_{j} y_{j}(t)} {\langle k \rangle} = \frac {k_{i}} {\langle k \rangle} Y(t), \label{step2}
\end{equation}
where we have defined the global quantity or field $Y(t)$, which no longer depends on the single neuron $i$.
Such a mean-field approximation is exact in the limit of $k_i \rightarrow\infty$ and it can be rigorously treated for massive graphs, where the degree $k_i$ scales with the number of neurons $N$. Furthermore, the HMF approximation allows to replace the dynamics for each single node with the dynamics of each class of nodes with degree $k_i$. 

In particular, in a network with both excitatory and inhibitory neurons, for each class of degree we introduce two variables $v_{k}^{E}$ and $v_{k}^{I}$ describing the membrane dynamics of excitatory and inhibitory neuron class respectively. Indeed, due to the facilitation mechanism,  the synaptic current reaching an inhibitory neuron is different from the current of an excitatory neuron and this gives rise to a different evolution of the membrane potential.
In this framework, four different heterogeneous mean fields need to be defined, i.e., the field produced and received by excitatory and inhibitory neurons respectively \cite{parma2016_HMF3inhib}. If we denote with $*, \dagger$ the indexes $E,I$ for excitation and inhibition, the four global fields are
\begin{equation}
\label{Y_defint}
Y_{\dagger, *}(t)\,=\,\int \frac {P_{*}(k)\, k\, y_{k}^{(\dagger,*)}} {\langle k \rangle}\,dk,
\end{equation}
where $y_{k}^{(\dagger,*)}$ represents the active resources of nodes with degree $k$ and of type $*$ outgoing towards a node of type $\dagger$. Similarly $x_{k}^{(\dagger,*)}$ and $z_{k}^{(\dagger,*)}$ are the available and inactive resources.

Accordingly, the synaptic fields $Y_{E}(t)$ and $Y_{I}(t)$ received respectively by excitatory and inhibitory neurons are
\begin{equation}
\label{Y_def}
\begin{split}
Y_{E}(t)&=+ ~ Y_{EE}(t)~ - ~Y_{EI}(t),\\
Y_{I}(t)&=+ ~ Y_{IE}(t) ~-~ Y_{II}(t),
\end{split}
\end{equation}
where the signs $+$ and $-$ take into account the excitatory and inhibitory nature of the presynaptic neuron. The final equations for the network dynamics in the HMF formulation are therefore
\begin{equation}
\label{HMF_end}
\begin{split}
\dot{v}_{k}^{\dagger}(t)&= a - v_{k}^{\dagger}(t) + \frac{g}{\langle k \rangle} k Y_{\dagger}(t), \\
\dot{y}_{k}^{(\dagger,*)}(t)&=  - \frac{y_{k}^{(\dagger,*)}(t)}{\tau_{in}} + u_{k}^{(\dagger,*)}(t) x_{k}^{(\dagger,*)}(t) S_{k}^{*}(t), \\
\dot{z}_{k}^{(\dagger,*)}(t)&= \frac{y_{k}^{(\dagger,*)}(t)}{\tau_{in}} - \frac{z_{k}^{(\dagger,*)}(t)}{\tau_{r}^{\dagger}}, \\
x_{k}^{(\dagger,*)}(t)& + y_{k}^{(\dagger,*)}(t) + z_{k}^{(\dagger,*)}(t) = 1, \\
u_{k}^{(E,*)}(t)&= U, \\
\dot{u}_{k}^{(I,*)}(t)&= - \frac{u_{k}^{(I,*)}(t)}{\tau_{f}} + U_{f} [1 - u_{k}^{(I,*)}(t)] S_{k}^{*}(t).
\end{split}
\end{equation}

Unless otherwise specified, the values of all parameters appearing in the equations are set according to accepted phenomenological values (Refs.~\cite{volman2005_values1, fuchs2009_values2}), after a proper time rescaling, as follows: $\tau_{in} = 0.2$, $\tau_{r}^{I} = 17\cdot \tau_{in}$, $\tau_{r}^{E} = 133\cdot \tau_{in}$, $\tau_{f} = 33.25$, $a = 1.3$, $g = 30$, $U_f = U = 0.5$. 

Integrating the set of Eqs.~(\ref{HMF_end}) through the same steps used for the dynamics (\ref{eqv})-(\ref{eqx}), a similar event-driven map can be obtained. In order to simulate the dynamics of the HMF model we need to perform a sampling of the $k$ values from the p.d.f $P(k)$. We point out that numerical simulations are now very efficient, since, through an importance sampling of the degree $k$, say we sample $M$ values of $k$, $M \sim \mathcal{O}(10^2)$  is sufficient to obtain simulations where the fluctuations due to discretization of the distribution are very small. In Sec.~\ref{sec6} we will compare the dynamics of finite large networks with the results of the HMF approach and we will show that two different behaviors emerge. In the quasi-periodic regime, at small $f_I$, the effect of a finite degree is simply a dynamical noise superimposed to the mean-field solution. More interestingly, we put into evidence that in the balance region the synchronous mean-field solution is unstable for the finite network and an asynchronous state is always reached asymptotically; however, in this case the lifetime of the synchronous state diverges with the connectivity, recovering also in this case the HMF dynamics for  large enough connectivities.

\section{\small SYNCHRONIZATION EFFECTS OF \\ INHIBITORY HUBS}\label{sec3}
\label{sec:sincro}
In this section we show numerical HMF results, putting into evidence the synchronization effects of inhibitory hubs.
As the hub nature of a neuron depends only on its degree, we need to choose suitable p.d.f.s $P_{E}(k)$ and $P_{I}(k)$. For simplicity we fix the two p.d.f.s to be both Gaussians with standard deviation $\sigma = 10$ and averages $\langle k_E \rangle$ and $\langle k_I \rangle$ for excitatory and inhibitory nodes, while the distance between the two distributions is $\Delta = \langle k_I \rangle - \langle k_E \rangle$. The larger $\Delta$ is, the more hub the inhibitory neurons are.
In particular, in our simulations we fix $\langle k_E \rangle = 100$ and we vary $f_I$ and $\Delta$, that tune the fraction of inhibitory nodes and how strong is their hub character, respectively.
From a technical ground, the continuous distributions $P_{E}(k)$ and $P_{I}(k)$ are efficiently sampled, obtaining $M$ discrete connectivity classes, each one including all neurons with degree $k \in (k_i, k_{i+1})$ (\textit{i} = 1, ..., $M-1$).

\begin{figure}
\subfigure[\label{fig:cfr_isim_tantef}]
{\includegraphics[width=1\columnwidth]{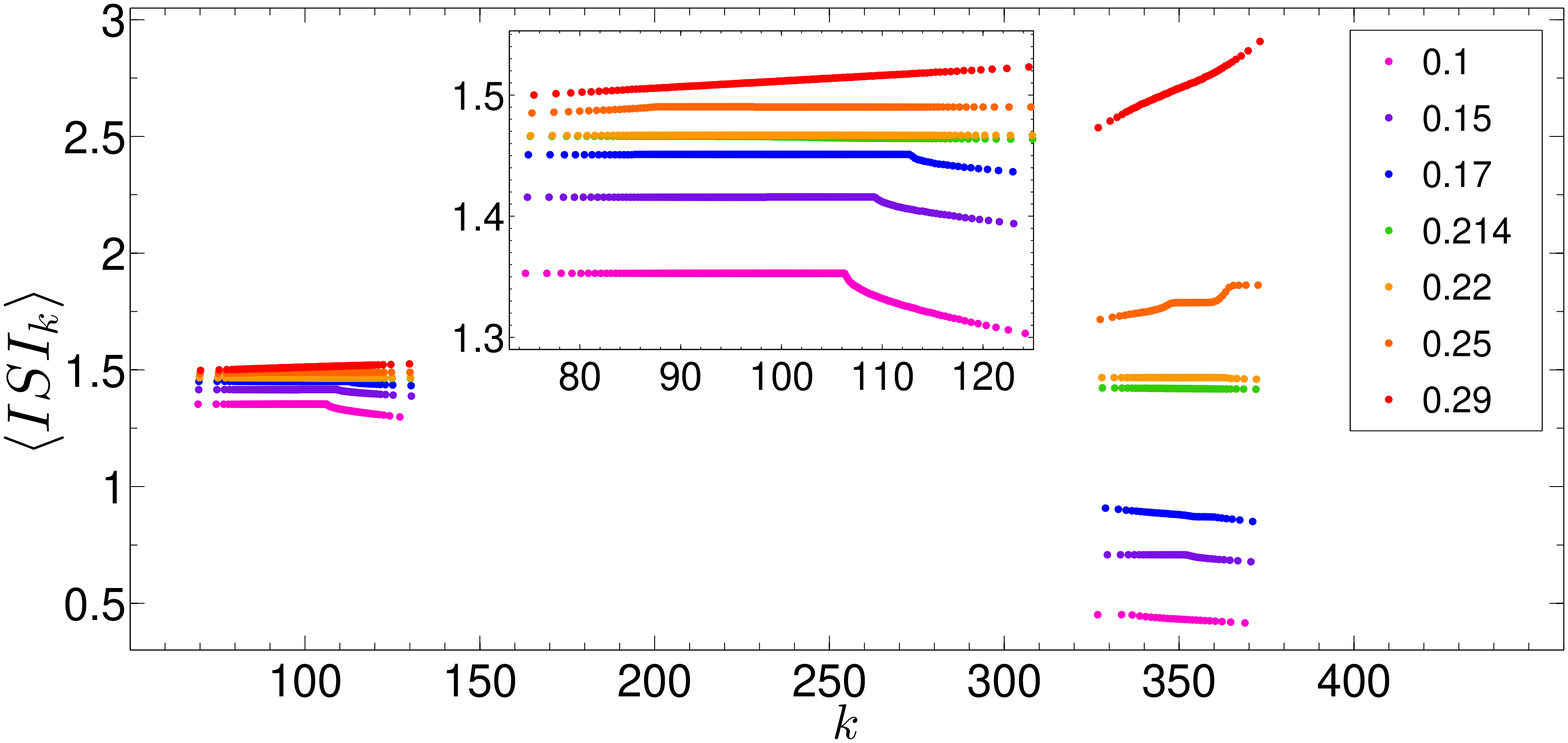}}
\subfigure[\label{fig:cfr_raster_tantef}]
{\includegraphics[width=1\columnwidth]{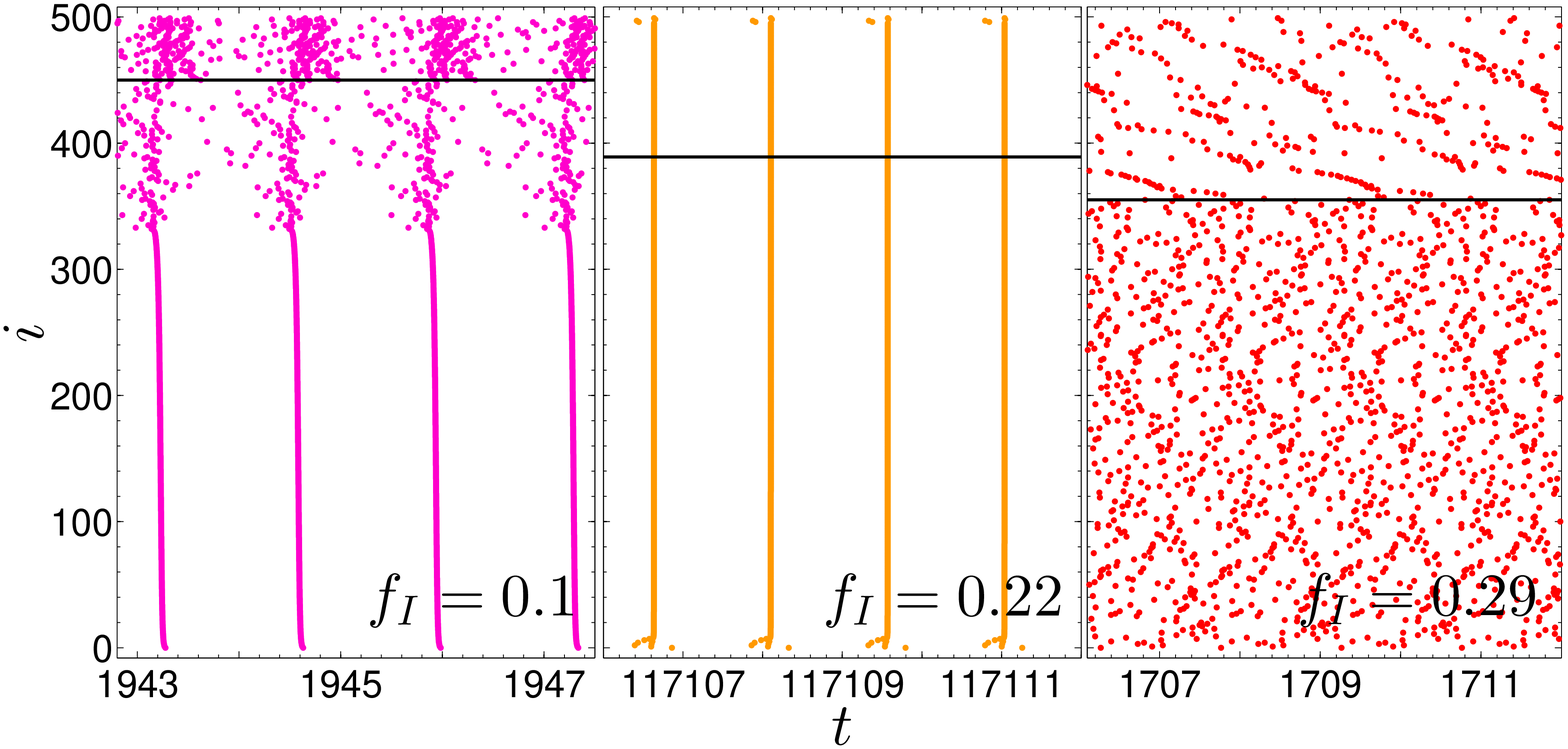}}
\caption{\footnotesize{(a) $\langle$ISI$_k\rangle$ as a function of the degree $k$ for networks with different inhibitory fractions $f_I$ (see the legend). $\Delta$ is fixed at $250$ and the number $M$ of connectivity classes is $500$. In the inset we show a zoom of the excitatory plateau. In panel (b) we display snapshots of the raster plots for different inhibitory fractions, to represent the three main dynamical regimes: partial synchronization for $f_I = 0.1$, total synchronization for $f_I = 0.22$, and an asynchronous state for $f_I = 0.29$. A dot in raster plots of HMF simulations represents the spike event of connectivity class $i$ at time $t$ and classes are ordered along the vertical axis according to their degree $k$. The degree classes below (above) the black line are excitatory (inhibitory). These plots refer to different dynamical regimes, however a time unit in the abscissa axis corresponds to about one global oscillation of the network (after this time interval all neurons have fired at least once).}}
\end{figure}

\begin{figure}
\centering
\includegraphics[width=1\columnwidth]{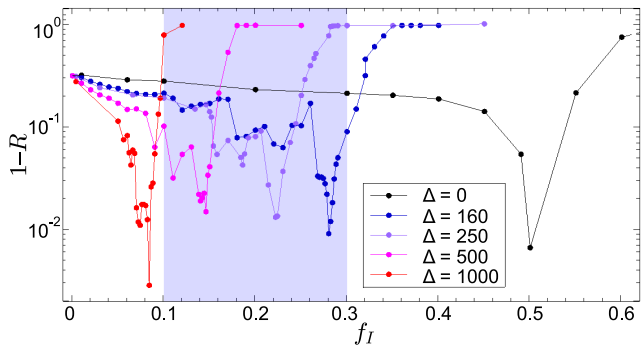}
\caption{\footnotesize{Network synchronization from the measure of $1 - R$ (where $R$ is the Kuramoto parameter) as a function of $f_I$ for different values of $\Delta$ (see the legend), in semilogarithmic scale. In all networks $M$ is equal to $500$. The closer to $0$ the parameter $1 - R$ is, the more synchronous the network. The grey-shaded area represents the interval of inhibitory fractions, which has been really observed in some brain tissues. Without hub nodes ($\Delta = 0$, black data), the network reaches a high synchronization level out of this region, differently from $160 \lesssim \Delta \lesssim 500$. Therefore, if we want to build synchronous networks with inhibitory fractions close to real values, the inhibitory neurons should be hubs, with an average degree 3-6 times greater than the excitatory average degree.}}
\label{fig:plot_kura_finale}
\end{figure}

First let us fix $\Delta = 250$ in order to observe the dynamical configurations of the network as a function of $f_I$. The raster plots in Fig.~\ref{fig:cfr_raster_tantef} have been obtained for three different values of this parameter. When $f_I$ is close to $0$ the network is partially synchronized and neurons are divided into two groups: excitatory neurons with low degree are locked, that is they evolve according a periodic dynamics and they share the same period, while all the others (excitatory nodes with higher degree and inhibitory hubs) are unlocked, displaying different periods and an aperiodic dynamics. The plot of $\langle$ISI$_k\rangle$ as a function of $k$, as shown in Fig.~\ref{fig:cfr_isim_tantef}, synthetically represent the dynamical properties of each connectivity class. Locked neurons, whose $\langle ISI_k \rangle$ does not depend on their degree, give rise to a plateau, as we can observe in that figure for $f_I = 0.1$ and $k < 106$. 

For larger values of $f_I$, the network synchronization grows. Indeed, the plateau in Fig.~\ref{fig:cfr_isim_tantef} in the excitatory region stretches, while a new inhibitory plateau appears. After the network has reached the total synchronization for a particular value of $f_I$ (which for $\Delta = 250$ is $0.214$), with all neurons locked and $\langle ISI_k \rangle$ belonging to two identical plateaus, the network becomes more and more asynchronous for larger $f_I$.

In order to quantify the synchronization level of the network, we calculate the Kuramoto parameter $R$ (Refs.~\cite{kuramoto2012_libro, parma2013_primo}), defined as follows:
\begin{equation}     
\label{kuradef}
R = \Biggl\langle \left | \frac{1}{M} \sum_{k=1}^{M} e^{2 \pi i [(t - t_{k}^{n})/({t_{k}^{n+1} - t_{k}^{n}})} \right | \Biggr\rangle,
\end{equation}
where the vertical bars represent the modulus of the complex number and the angle brackets the temporal average value of that quantity, while the exponent is the phase we assign to the connectivity class $k$. The closer to $0$ ($1$) $R$ becomes, the more asynchronous (synchronous) the dynamics is. $M$ is the total number of classes, $t_{k}^{n}$ is the time of the \textit{n}th impulse of class $k$, and $t_{k}^{n+1}$ is the time of the following $(n+1)$th impulse.

For each $\Delta$, we vary $f_I$ from $0$ to a value that corresponds to the totally asynchronous network and we compute the relative Kuramoto parameter $R$, namely the parameter $1 - R$, plotted in Fig.~\ref{fig:plot_kura_finale} as a function of $f_I$ in semilogarithmic scale to make the transition from the asynchronous to the synchronous regime more evident. For all values of $\Delta\geq 0$, $R$ first increases towards $1$, meaning that, at small values of $f_I$, increasing the fraction of inhibitory neurons really improves the synchronization of the whole network. Then, the synchronization level reaches a peak, where the network is totally synchronized and $R$ is very close to $1$. We notice that for different values of $\Delta$ this peak corresponds to a different and specific value for the inhibitory fraction. Finally, $R$ quickly decays towards $0$ and the network becomes asynchronous. 

If inhibitory neurons do not feature a higher connectivity with respect to the excitatory ones ($\Delta = 0$), we observe that the inhibitory-driven synchronous regime is observed only for high values of $f_I$. Nevertheless, experimental observations suggest that the fraction of inhibitory neurons is much smaller ($f_I \sim 10-30 \%$) \cite{fractinh1, fractinh2}. Actually, we show how by increasing the hub character of inhibitory neurons one can observe such a synchronized regime for much smaller values of $f_I$, reaching phenomenological values (grey-shaded region in Fig.~\ref{fig:plot_kura_finale}) for $2 \times 10^2 \lesssim \Delta \lesssim 10^3$. 
Apart from the quantitative reliability of our results with respect to the experimentally suggested fraction of inhibitory neurons, Fig.~\ref{fig:plot_kura_finale} shows how the hub character of inhibitory neurons is a fundamental ingredient to allow a small fraction of inhibitory nodes to drive the synchronization of the overall network, thus underlining the importance of inhibitory neurons connectivity in affecting network synchronization.
The narrow region of $f_I$ where the network is highly synchronized is a consequence of a dynamical balance between the excitatory and the inhibitory mechanisms, which will be discussed in the next section. 

\section{\small BALANCE REGIME}
\label{sec4}
In the totally synchronous phase, all neurons follow the same dynamics and the evolution of $v_k^\dagger(t)$ no longer depends on $k$. This means that the last term in the first equation in (\ref{HMF_end}) should be $0$, i.e., the excitatory and the inhibitory contributions in the definition of $Y_{\dagger}(t)$ cancel each other out.
Writing $Y_{\dagger}(t)$ according to Eqs.~(\ref{Y_def}) and (\ref{Y_defint}), we obtain
\begin{equation}
\begin{split}
Y_{\dagger,E}(t) - Y_{\dagger,I}(t) &= 0, \\
\int \frac {{P_{E}(k)} k y_{k}^{(\dagger,*)}} {\langle k \rangle} dk &= \int \frac {{P_{I}(k)} k y_{k}^{(\dagger,*)}} {\langle k \rangle} dk,\\
\frac{y}{\langle k \rangle} \int P_{E}(k) k dk &= \frac{y}{\langle k \rangle} \int P_{I}(k) k dk,\\
\langle k_E \rangle f_{E} &= \langle k_I \rangle f_{I}, \\
\langle k_{E} \rangle (1 - f_{I}) &= (\Delta + \langle k_{E} \rangle) f_{I},
\end{split}
\label{eq15}
\end{equation}
where in the second step we have used $y_{k}^{(\dagger,*)} = y\, (\forall k)$ and we factorized them out of the integrals, since the network is totally synchronous and all microscopic variables are equal. Simplifying $y / \langle k \rangle$ from the third equation, we obtain that the left integral represents the average value of $P_{E}(k)$ multiplied by $f_{E}$ and similarly the right integral [see Eq.~(\ref{f_I})]. Solving the last equation for $f_I$ we get the inhibitory fraction $f_{I}^{B}$ for the balance regime. We verified indeed that, by fixing the inhibitory fraction to the predicted balance value, the Kuramoto parameter is maximum and very close to $1$: $R = 0.987-0.997$ for the different values of $\Delta$, as shown in Fig.~\ref{fig:plot_kura_finale}. Furthermore, in the balance regime all neurons are locked and characterized by the same period, that can be computed from the membrane potential equation [first Eq.~in (\ref{HMF_end})] without the field term. The period of these free oscillators is simply $\langle$ISI$_k\rangle = \ln [a / (a-1)] = 1.466\,(\forall k)$. 

Out of this balance phase, one of the two mechanics, excitatory or inhibitory, exceeds the other. We measure the relative weight of the excitatory and the inhibitory component introducing the following quantities: 
\begin{equation}
\begin{split}
W_E = \left \langle \frac{Y_{EE}(t) - Y_{EI}(t)}{Y_{EE}(t) + Y_{EI}(t)} \right \rangle, \\
W_I = \left \langle \frac{Y_{IE}(t) - Y_{II}(t)}{Y_{IE}(t) + Y_{II}(t)} \right \rangle,
\end{split}
\end{equation}
where the angle brackets represent again the temporal average. $W_E$ refers to the field received by an excitatory neuron, while $W_I$ refers to the field received by an inhibitory neuron. If $W_{E} \simeq W_{I} \simeq 1$, the dynamics of the whole network is purely excitatory and inhibitory neurons do not substantially contribute to the global fields. In the limit of $W_{E} \simeq W_{I} \simeq -1$ the dynamics is totally driven by the inhibitory mechanism, even though this limit is theoretical, since a purely inhibitory network is not in a spiking regime. In Fig.~\ref{fig:plot_EvsI} we plot $W_{E}$ and $W_{I}$ for the values of $\Delta$ and $f_I$ considered in Fig.~\ref{fig:plot_kura_finale}.
We immediately observe that for the balance value $f_I^B$ of the inhibitory fraction we get $W_{E} = W_{I} = 0$.

\begin{figure}
\includegraphics[width=1\columnwidth] {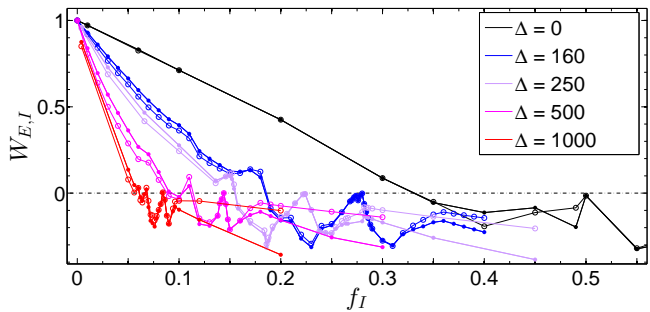}
\caption{\footnotesize{Relative weights of the excitatory and the inhibitory component in driving the networks dynamics with $M = 500$ connectivity classes, from the measure of $W_E$ and $W_I$ for different values of $\Delta$ (see the legend) and $f_I$. The balance value $f_I^B$ of the inhibitory fraction for which $W_{E} = W_{I} = 0$ is $0.5$ for $\Delta = 0$, $0.28$ for $\Delta = 160$, $0.22$ for $\Delta = 250$, $0.143$ for $\Delta = 500$ and $0.083$ for $\Delta = 1000$.}}
\label{fig:plot_EvsI}
\end{figure}

However, we see that the increase of $f_I$ does not necessarily lead to a monotonous increase in the inhibitory contribution, but some unexpected oscillations appear, due to nontrivial equilibrium mechanisms between the excitatory and the inhibitory dynamics outside the balance region. 

The Kuramoto plot in Fig.~\ref{fig:plot_kura_finale} can be now compared with what emerges in Fig.~\ref{fig:plot_EvsI}. At first, for each $\Delta$, as the fraction of inhibitory hubs increases and the network dynamics becomes more inhibitory (in Fig.~\ref{fig:plot_EvsI} $W_E$ and $W_I$ are indeed decreasing), the synchronization improves, starting from a partially synchronous evolution, until the totally synchronous state is reached for the balance value of $f_I$. Then, the further increase of $f_I$ leads to the prevalence of the inhibitory dynamics, since both $W_E$ and $W_I$ become negative, and it gives rise to asynchronous regimes.
By increasing $\Delta$, this dynamical transition from a synchronous to an asynchronous configuration occurs in a narrower interval of $f_I$.

\begin{figure}
\centering
\includegraphics[width=1\columnwidth]{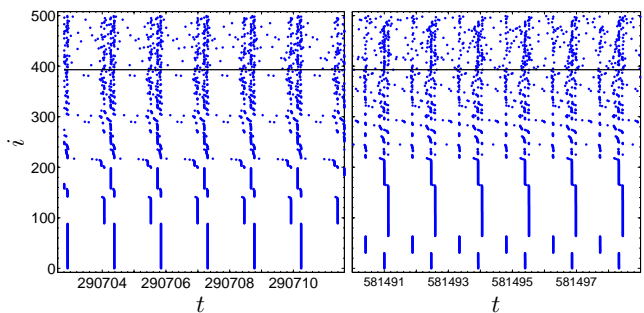}
\caption{\footnotesize{Raster plots of a metastable partially synchronous state for the network with $M = 500$, $\Delta = 250$, and $f_I = 0.214$, starting from random initial conditions. These plots are two snapshots, taken at different times during the same dynamical evolution. The degree classes below (above) the black line are excitatory (inhibitory).}}
\label{fig:raster_meta}
\end{figure}

When $\Delta$ has intermediate values, i.e., $150 < \Delta < 500$, the region immediately before the balance (for example $0.2 < f_I < 0.214$ for $\Delta = 250$) is characterized by a highly metastable regime of partial synchronization, where almost all excitatory neurons are locked with very similar frequencies but different phases and therefore they split into different synchronized groups, as we can see in the raster plots in Fig.~\ref{fig:raster_meta}. In this regime, inhibitory neurons are unlocked, even if they spike almost simultaneously with one of the excitatory locked groups, but the group changes in time. Moreover, the system is metastable. Indeed, even the configuration of the synchronous excitatory groups changes on very long-time scales, typically of thousands of oscillations; for example the raster plots in Fig.~\ref{fig:raster_meta} are different snapshots taken during an evolution of the same network. As one can expect, the structure of these metastable states depends on the initial conditions of the microscopic variables. In particular, if one imposes synchronous initial conditions (potentials and synaptic resources are equal for all neurons), the dynamics reaches a configuration where a single group of synchronous neurons is present and still some are unlocked. This configuration seems to be the asymptotic stable situation since, at variance with Fig.~\ref{fig:raster_meta}, we do not observe a system metastability on long-time scales. However, we never reach such a total synchronous state when starting from random initial conditions, an evidence of the slow dynamics.
For this reason, we point out that in Fig.~\ref{fig:plot_kura_finale} the value of the Kuramoto parameter of networks with this metastable dynamics has been computed referring to the total synchronous state, namely to the state the network assumes using total synchronized initial conditions. The robustness of the metastable dynamics is also preserved, even when we modify the discretization for the degree classes from the initial continuous p.d.f.~$P(k)$, which is another relevant parameter for the simulations.

\section{\small RESPONSE TO AN EXTERNAL STIMULUS}
\label{sec5}
In this section we investigate the input processing features of the network by analyzing the reaction of our system to an external stimulus. In particular, we focus on the transient dynamics at the stimulus offset as the space trajectory of the system in this transient has been supposed to be crucial for the stimulus detection \cite{timme}. In particular, the longer the time to return to the unperturbed state is, the better the system is able to detect and distinguish the input \cite{odorstimuli}.

We will show that the metastable states, like those in Fig.~\ref{fig:raster_meta}, can be affected by the external perturbation longer than the other dynamical regimes. Indeed, as the configuration of the synchronous excitatory groups depends on the initial conditions and changes on long-time scales, an external synchronous stimulus, creating a new group, can produce an alteration of the network dynamics for long-time intervals.

\begin{figure}
\centering
\includegraphics[width=1\columnwidth]{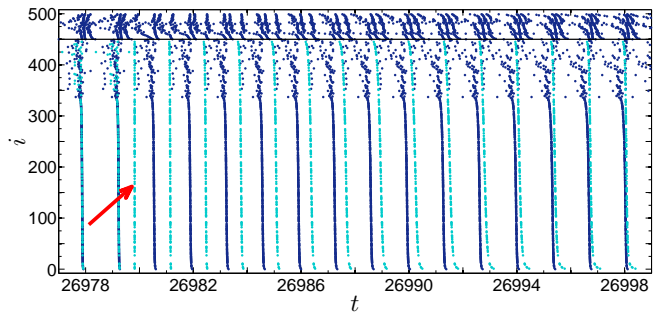}
\caption{\footnotesize{Raster plot of a network with $M = 500$ connectivity classes, $\Delta = 250$ and $f_I = 0.1$. At a given time (see the red arrow), we apply a synchronizing external stimulus to $30\%$ of excitatory neurons, randomly chosen. After about ten global oscillations, the stimulated nodes recover the initial dynamics. The degree classes below (above) the black line are excitatory (inhibitory). Light (dark) blue dots represent the spike events of stimulated (not stimulated) nodes.}}
\label{fig:raster_stimolo}
\end{figure}

We consider $\Delta = 250$ and vary $f_I$, so that the different dynamical regimes can be explored. We stimulate only the excitatory nodes, since they are the most abundant in real networks, as it has been experimentally observed, and therefore they are more likely to be affected by external perturbations with respect to inhibitory cells. In all networks, we randomly choose $30\%$ of the excitatory neurons and we synchronize them artificially; e.g., we can force the simultaneous firing of the stimulated neurons, when the excitatory field $Y_{E}(t)$ reaches its minimum value (see Fig.~\ref{fig:raster_stimolo}). Such forced synchronization, that may represent external visual or odor input, could encode the intensity of the stimuli: the higher the number of synchronous units the higher the release of neurotransmitter $y$ due to the spiking of forced neurons. After the perturbation, we compute the reduced Kuramoto parameter $R(t)$ of the neurons which has been initially boosted to $1$ by the input. Then we measure the time required to return to the original value and therefore how long it takes for the network to absorb the external stimulus and come back to the original dynamical scenario. In Fig.~\ref{fig:plot_kura_stimolo} we plot the evolution of the parameter $1 - R(t)$ as a function of time, for different $f_I$. At the beginning, $1 - R(t)$ is close to $0$, because the stimulus roughly improves the phase coherence of the stimulated group; then it increases, in a fast or slow way depending on the inhibitory fraction which is present, and it returns to the original value, represented by the plateau on the right of Fig.~\ref{fig:plot_kura_stimolo}. 

\begin{figure}
\centering
\includegraphics[width=1\columnwidth]{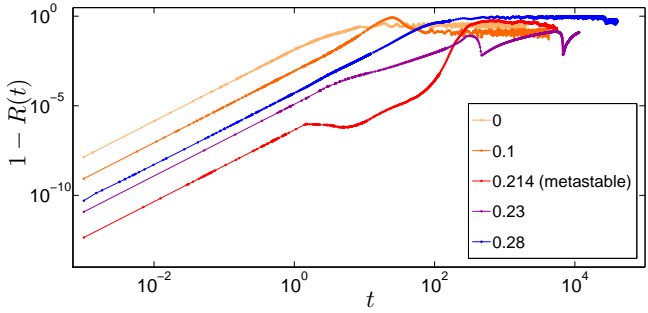}
\caption{\footnotesize{Transient dynamics at the stimulus offset: return to the original synchronization level after the application of an external input. We plot the parameter $1 - R(t)$ of the stimulated nodes as a function of time, for networks with $M = 500$, $\Delta = 250$, and different values of $f_I$ (see the legend). The external synchronous stimulus is always applied at $t = 0$ to $30\%$ of excitatory neurons, randomly chosen.}}
\label{fig:plot_kura_stimolo}
\end{figure}

We observe that, when the inhibitory fraction is too small or too high, the network quickly destroys the new synchronization imposed by the stimulus and after about ten oscillations the initial state is recovered. In the network with the metastable dynamics, the duration of the perturbation is $20$ times longer. Therefore, in a partially synchronized configuration, the metastable dynamics of the network stores information of a synchronous perturbation for long-time scales. 

In Fig.~\ref{fig:plot_kura_stimolo}, the initial growth of the reduced Kuramoto parameter follows a simple power law, namely $1 - R(t)\sim t^2$, which can be explained by expanding $R(t)$ for small $t$: 
\begin{equation}
\label{kura_exp}
\begin{split}
     R(t) &= \Biggl | \frac{1}{M} \sum_{k=1}^{M} e^{i \frac{t - t_{k}^{n}}{t_{k}^{n+1} - t_{k}^{n}}} \Biggr | \\
     &\approx \Biggl | \frac{1}{M} \sum_{k=1}^{M} \Bigl(1 + i \frac{t}{t_{k}^{n+1}} - \frac{1}{2} \frac{t^2}{(t_{k}^{n+1})^2} \Bigr ) \Biggr | \\ 
     &\approx \sqrt{1 - \frac{t^2}{M} \Bigl(\sum \frac{1}{(t_{k}^{n+1})^2} \Bigr) + \frac{t^2}{M^2} \Bigl(\sum \frac{1}{t_{k}^{n+1}} \Bigr)^2} \\
     &\approx 1 - \frac{t^2}{2} \Bigl( \Bigl\langle \frac{1}{(t_{k}^{n+1})^2} \Bigr\rangle - \Bigl\langle \frac{1}{t_{k}^{n+1}}\Bigr\rangle^2 \Bigr), 
\end{split}
\end{equation}
where the sum runs only on the nodes which have received the synchronous stimulus and therefore show $t_{k}^{n} = 0$. 
 
Similar plots can be obtained also for different connectivities of the inhibitory hubs, (we made explicit checks for  $\Delta = 160$ and $\Delta = 500$): the prebalance regime is again characterized by partially synchronous metastable states and the time required to return to the original synchronization level is at least an order of magnitude longer than in other regimes, as in Fig.~\ref{fig:plot_kura_stimolo}.

\begin{figure}
	\centering
	\subfigure
	{\includegraphics[width=1\columnwidth]{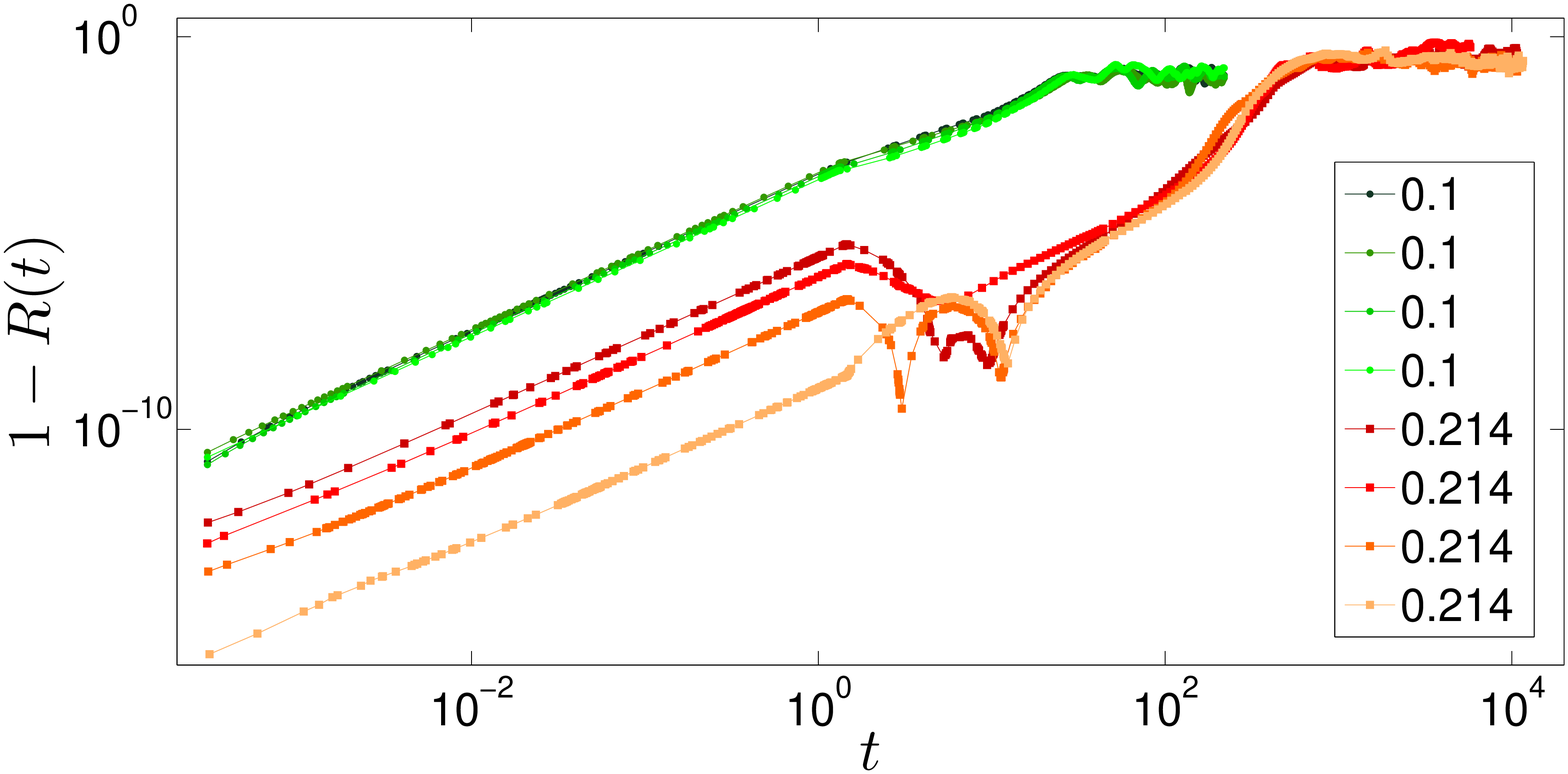}}
	\subfigure
	{\includegraphics[width=1\columnwidth]{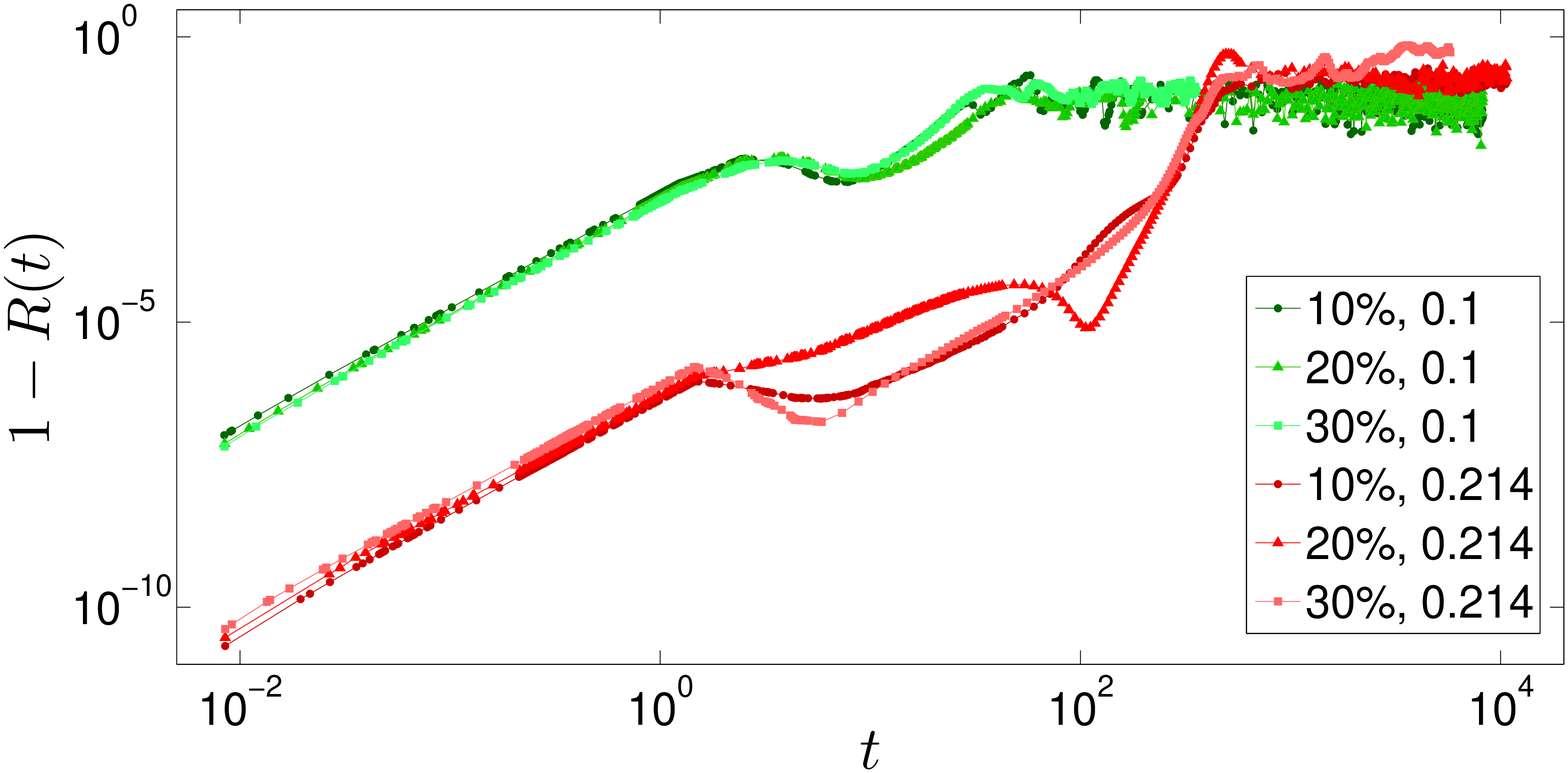}}
	\caption{\footnotesize{Response of the network to external stimuli applied at different times and with different intensity. In both figures, we plot the parameter $1 - R(t)$ of the stimulated nodes as a function of time and we consider networks with $M = 500$ and $\Delta = 250$. Green (red) shades data refer to $f_I = 0.1 (0.214)$. Upper panel: for each network, we choose four different times to switch on a stimulus with the same intensity, i.e., on $30\%$ of excitatory neurons, randomly chosen. All data are shifted so that the onset of the perturbation is at $t = 0$. Lower panel: we compare the transient dynamics for stimuli applied on $10\%$, $20\%$, and $30\%$ of the excitatory neurons at $t = 0$.}}
	\label{fig:robust_stim}
\end{figure}

We verify the robustness of our results showing that the response of the network to a given stimulus is independent of the specific time the perturbation is switched on. In the upper panel of Fig.~\ref{fig:robust_stim} we randomly choose four different times to apply the same stimulus to the same network, so now the perturbation does not necessarily occur at the minimum of the excitatory fields $Y_E(t)$. In all trials we obtain similar transient dynamics and the relaxation times are consistent.

\begin{figure*}
\includegraphics[width=2\columnwidth]{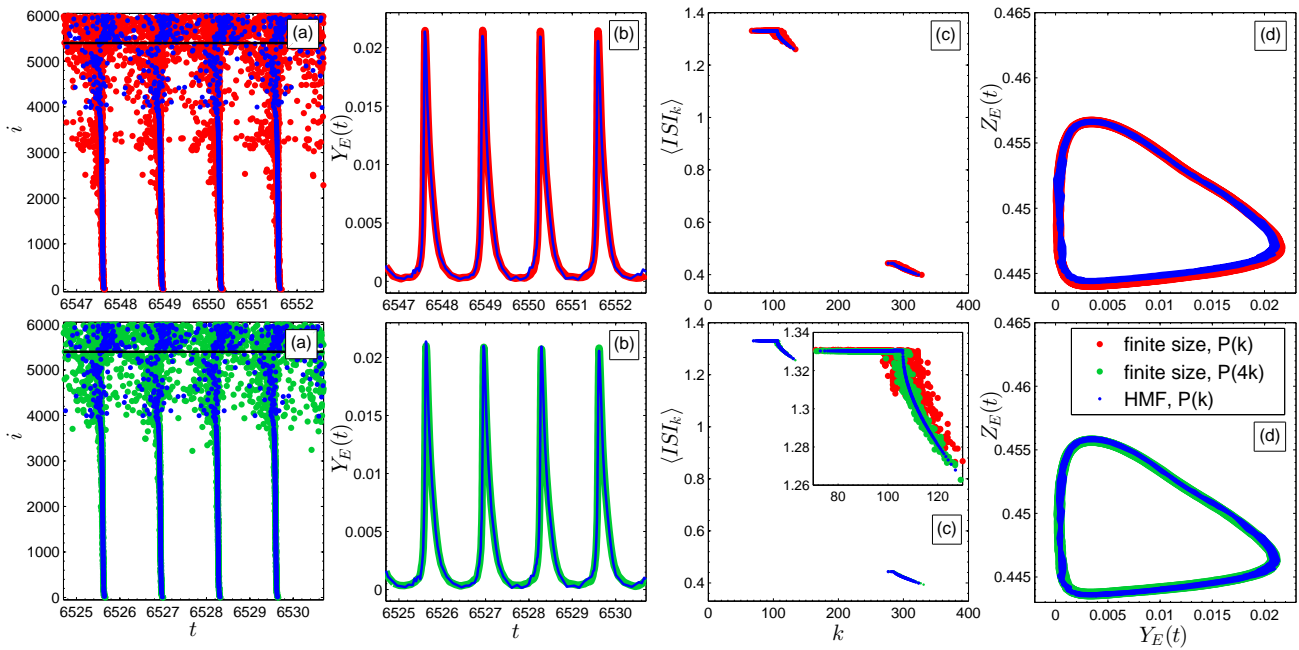}
\caption{\footnotesize{Comparison of the dynamics of a network with $f_I = 0.1$, simulated through both the finite size and the HMF approach. In the upper figures red data are produced evolving the finite size equations with $N = 6000$ neurons, and blue data the HMF ones with $M = 500$ connectivity classes. The p.d.f.~is characterized in both cases by $\langle k_E \rangle = 100$, $\Delta = 200$, and $\sigma = 10$. In the lower figures we show how the HMF approximation improves as we approach the limit of $k_i \rightarrow \infty$. Again, green (blue) data evolve according to the finite size (HMF) approach, but for the finite size network we have quadrupled the degrees (new p.d.f.~parameters: $\langle k_E \rangle = 400$, $\Delta = 800$, $\sigma = 40$). Thanks to this higher connectivity, the finite size dynamics better converges to the same HMF network. In panels (a) we compare the raster plots. As in the HMF figures $M$ is equal to $500$, the HMF indices have been opportunely rescaled for a better overlap with the red and green data. Data below (above) the black line refer to excitatory (inhibitory) nodes. (b) Field $Y_E(t)$ as a function of time. The comparison of $Y_I(t)$ gives similar results. (c) $\langle$ISI$_k\rangle$ as a function of the degree $k$. In the lower panel, we divided the degrees of green data by $4$, to overlap finite size data with HMF points. The inset is a zoom of the $\langle$ISI$_k\rangle$ only for the excitatory neurons and for all three sets of data. (d) Attractors $Y_{E}(t)$ vs $Z_{E}(t)$.}}
\label{fig:cfr_hmffs_f1000}
\end{figure*}

The same qualitative results are also obtained considering different fractions of excitatory neurons, e.g., $10\%$ or $20\%$, (lower panel of Fig.~\ref{fig:robust_stim}). We exclude stimuli larger than $50\%$, as these cannot be considered a perturbation of the system dynamics. Though different stimuli will generally correspond to different responses of the system, as the release of neurotransmitters $y$ produced by the external input depends on the intensity of the input itself, the quantity $1 - R(t)$ refers only to the perturbed neurons so that we can compare data obtained in different stimulations.  We observe that, independently of the input intensity, the metastable states are always affected by the external perturbation for times that are an order of magnitude longer than the other dynamical regimes of partial synchronization.

Let us point out that an intriguing result of this analysis is that, in order for the system to work in such an efficient regime, the network should be closed to the balance state where the connectivity ratio between inhibitory and excitatory neurons is equal to the inverse of the corresponding fractions [see Eq. (\ref{eq15})]. This result depends on the nature of the model and on the specific distribution of network connections but nonetheless it is an interesting prediction that might be tested experimentally.

\section{\small BEYOND THE HMF APPROXIMATION}
\label{sec6}
In the last part of this work, we compare the HMF approximation results with the dynamics on a finite network. Although a series of papers have already addressed this issue, we will show that in the balance regime the result is not trivial and further discussions are required. Accordingly, we compare the HMF approach, given by Eqs.~(\ref{HMF_end}), with the simulation of the Eqs.~(\ref{eqv})-(\ref{sincurr}), where the size is $N$ and the degree $k_i$ is extracted from $P_{E}(k)$ and $P_{I}(k)$. Clearly we expect the result to improve for large connectivities of the finite graph.
This can be verified by considering networks with different average degrees, which are approximated by the same HMF equations. In particular, we introduce a class of random graphs, where $P_{E}(k)$ and $P_{I}(k)$ are both Gaussian distributions with averages $\langle k_E \rangle$ and $\langle k_I \rangle = 3 \langle k_E \rangle$ ($\Delta = 2 \langle k_E \rangle$) and with standard deviation $\sigma = \langle k_E \rangle/10 $ for both excitatory and inhibitory neurons. By varying the values of $\langle k_E \rangle$ we get finite graphs with different average connectivity, which are approximated by the same HMF equations. Finally, fixing the fraction $f_I$, we can consider the various dynamical regimes: quasi-periodic, synchronous with balance, and asynchronous.

In Fig.~\ref{fig:cfr_hmffs_f1000} we consider $f_I = 0.1 $, i.e., the regime of partial synchronization, when $f_I$ is (much) smaller than its balance value $f_{I}^{B}$.
The blue data are obtained from the simulation of the HMF equations, while the upper panels refer to $\langle k_E \rangle=100$ and the lower panels to $\langle k_E \rangle=400$. Panels (a) represent the raster plots and panels (b) the synaptic fields $Y^{E}(t)$, which in the finite size approach are defined as
\begin{equation}
	\label{Y_def_FS}
	Y_{E}(t) = + Y_{EE}(t) - Y_{EI}(t)\,=\frac{1}{L} \Big(\sum_{i,j \in E,E} y_{ij} k_{i} - \sum_{i,j \in E,I} y_{ij} k_{i} \Big),
\end{equation}
where $L$ is the total number of links, $\sum_{i,j \in E,I}$ means that the sum is restricted to the $y$ resources from an inhibitory node to an excitatory one and similarly for $\sum_{i,j \in E,E}$. Panels (c) show the time average $\langle$ISI$_k\rangle$ of interspike interval as a function of the connectivity and finally in panels (d) we plot the microscopic attractors $Y_{E}(t)$ vs $Z_{E}(t)$ [in the finite size graph $Z_{E}(t)$ is defined analogously to Eq.~(\ref{Y_def_FS}), using the resources $z_{ij}$ instead of $y_{ij}$, while in the HMF simulation it is computed similarly to Eqs.~(\ref{Y_defint}) and (\ref{Y_def}), using $z_{k}^{(\dagger,*)}$ instead of $y_{k}^{(\dagger,*)}$]. 

Figure \ref{fig:cfr_hmffs_f1000} shows that HMF and finite size networks display a similar behavior and the main effect of finite connectivity is to superimpose on the HMF dynamics a noise, which is clearly vanishing by increasing the connectivity of the finite size graph. Analogous conclusions can be found for all the dynamical regimes where the fraction of inhibitory neurons is much smaller than the balance value.
A similar accordance can be observed also in the asynchronous regime, when $f_I > f_{I}^{B}$. As the network activity is now aperiodic and the fluctuations are dominant both for the synaptic fields and for the attractors, we do not show the comparison plots.

\begin{figure}
	\centering
	\subfigure
	{\includegraphics[width=1\columnwidth]{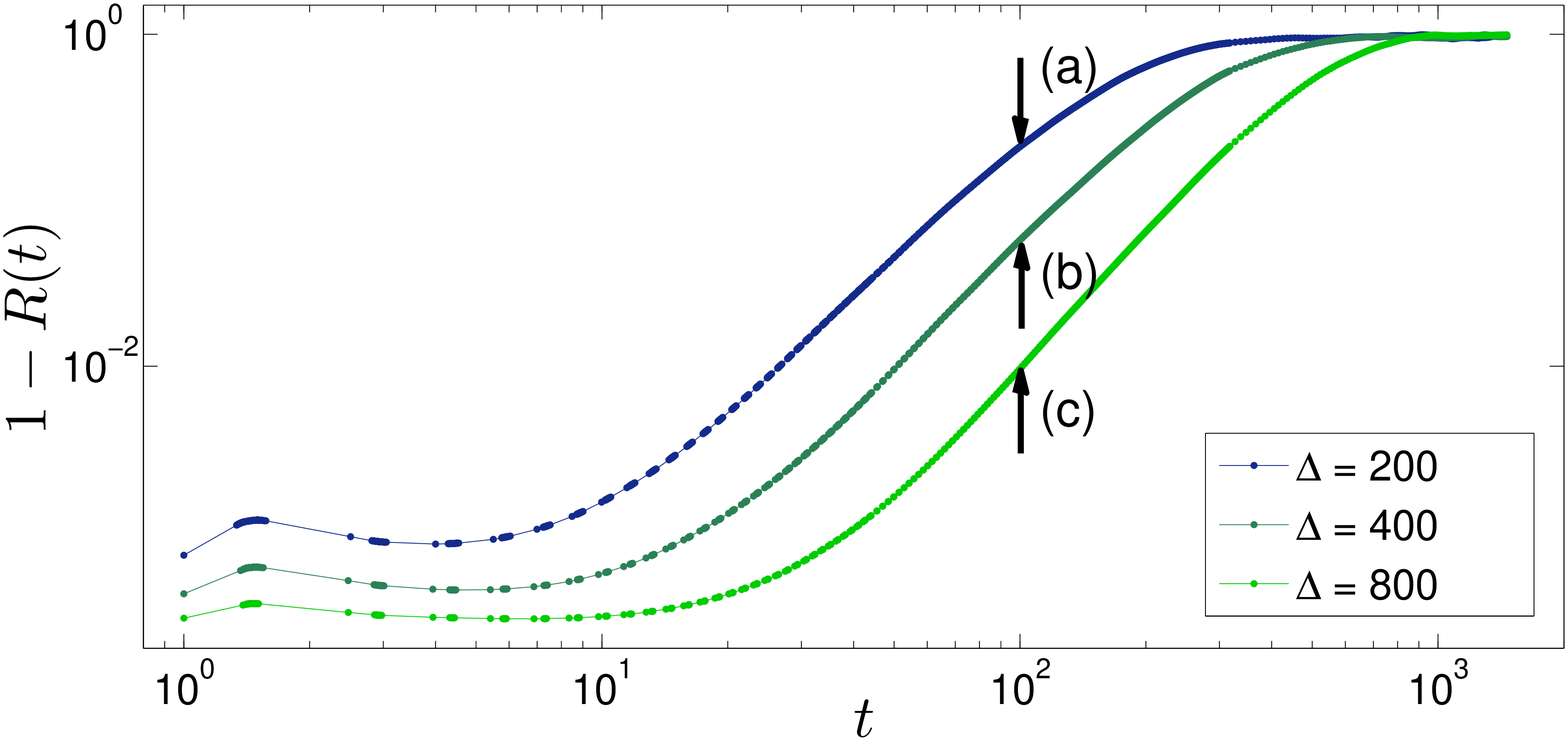}}
	\subfigure
	{\includegraphics[width=1\columnwidth]{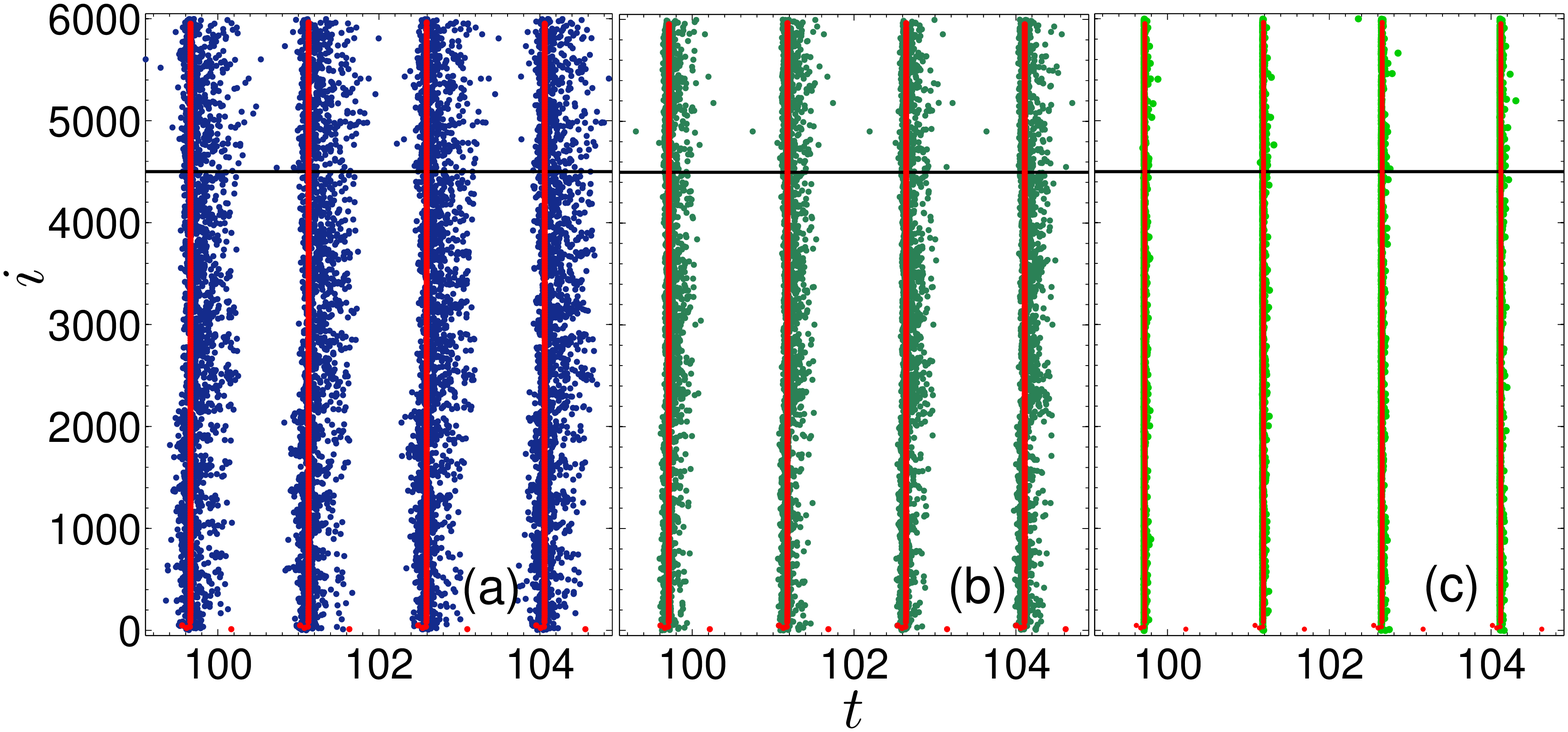}}
	\caption{\footnotesize{Stability of the synchronous state in finite size simulations. In the upper panel we compare the time evolution of the parameter $1 - R(t)$ (in logarithmic scale) in three different networks with $N = 6000$ neurons and $f_I = 0.25$. The p.d.f.~of the first network (blue data) is characterized by $\Delta = 200$, $\langle k_E \rangle = 100$, $\sigma = 10$; the p.d.f.~of the second network is doubled (dark green data, $\Delta = 400$, $\langle k_E \rangle = 200$, $\sigma = 20$); the p.d.f.~of the last network is quadrupled (light green data, $\Delta = 800$, $\langle k_E \rangle = 400$, $\sigma = 40$). The external stimulus is applied at time $t = 0$ to all neurons. At the beginning $1 - R(t)$ is almost null, because the system is totally synchronous, but then due to the finite size fluctuations it returns to $1$, that is the network evolves towards its stable asynchronous state. In correspondence of the black arrows, we show in the three lower figures the microscopic dynamics of the networks through their raster plots. The indices below (above) the black line refer to excitatory (inhibitory) neurons.}}
	\label{fig:stab_synchr_state}
\end{figure}

A substantial difference emerges in the balance regime, where the HMF approach is characterized by a totally synchronous dynamics. Indeed, in this case the asymptotic stable state for the finite size system seems to be asynchronous. In particular, the synchronization in the mean-field formulation is a consequence of a perfect field subtraction, as we have seen in Sec.~\ref{sec4}, but in networks with finite connectivity fluctuations dominate and destroy the phases locking inducing the synchronization.
However, the totally synchronous state characterizing the HMF equations emerges in the finite size system as a metastable state, whose lifetime increases with the network connectivity. 
In Fig.~\ref{fig:stab_synchr_state} we consider the balance regime $f_I = f_{I}^{B}=0.25$ for $\Delta = 200$: the lower panel shows the raster plots of three different networks, taken from time $t = 100$ after that a stimulus synchronized the whole dynamics. In the first network, with lower connectivity, the synchronous configuration is for the most part destroyed and the dynamics is quickly returning to the asynchronous state, but in the other raster plots, where the degree is two or four times larger, the network still remains quite synchronous, showing that the lifetime of the periodic dynamics diverges with the connectivity. 
The synchronization of the system can be measured using the parameter $1 - R(t)$ and plotting it as a function of time (the totally synchronous stimulus is applied at $t = 0$). The behavior of $1 - R(t)$ is illustrated in the upper panel of Fig.~\ref{fig:stab_synchr_state}, underlining that the larger the connectivity in the network is, the more stable the synchronous state is.

\begin{figure}
	\centering
	\subfigure
	{\includegraphics[width=1\columnwidth]{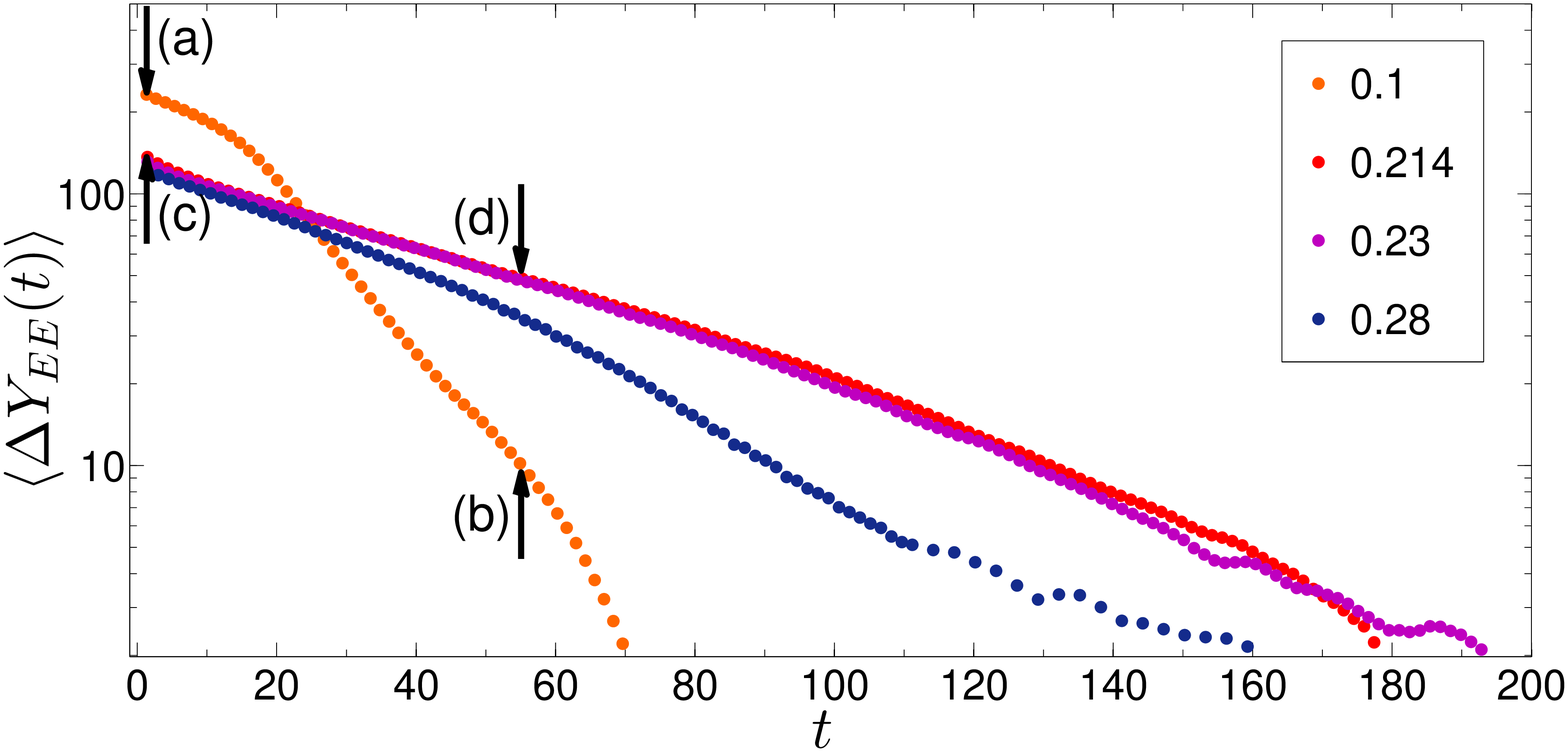}}
	\subfigure
	{\includegraphics[width=1\columnwidth]
	{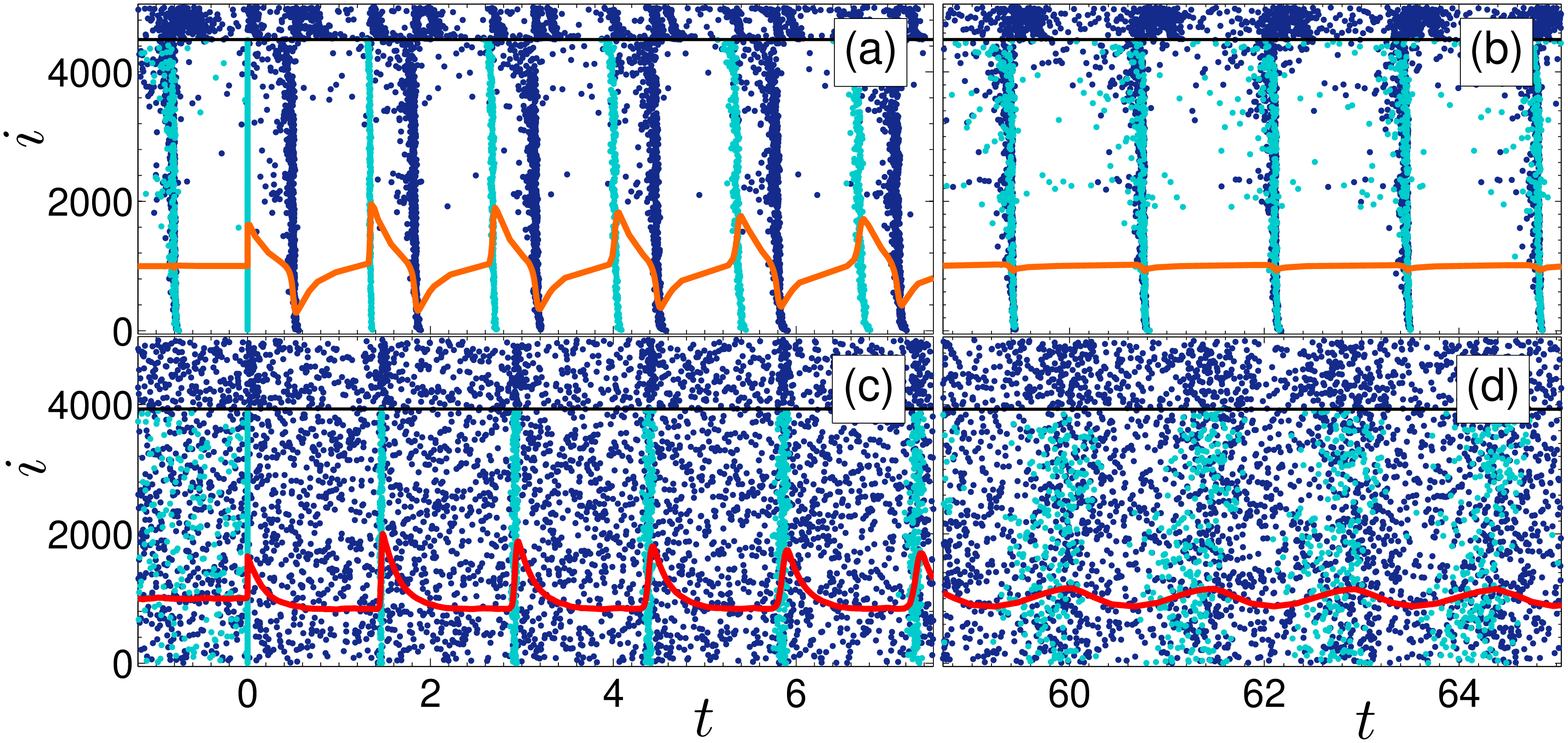}}
	\caption{\footnotesize{Dynamics of finite size networks after the application of an external stimulus at time $t = 0$ to $30\%$ of excitatory neurons, randomly chosen, with $N = 5000$ neurons, $\Delta = 250$, $\langle k_E \rangle = 100$, $\sigma = 10$ and different inhibitory fractions $f_I$ (see the legend). For each network we plot in the upper panel (in semilogarithmic scale) the temporal average values $\langle \Delta Y_{EE}(t) \rangle$ as a function of time, starting from $t = 0$. In correspondence of the black arrows, in the lower panels we show snapshots of the raster plot taken at different times for the network with $f_I = 0.1$ [(a) and (b)] and with $f_I = 0.214$ [(c) and (d)]. The spike events of stimulated (nonstimulated) nodes are represented by light (dark) blue dots, while the orange and red lines are the signals defined in Eq.~(\ref{form:stim_signal}). To better overlap signals with raster plots, we shifted them adding a positive constant: this is why they are not null before the stimulus. After the stimulus, signals turn on sharply and then decay. The indices below (above) the black line refer to excitatory (inhibitory) neurons.}}
	\label{fig:stim_fs}
\end{figure}

Eventually, we show that, even if the HMF and finite size dynamics have different attractors, the conclusions we have drawn regarding the high performance for input detection in this dynamical regime are still valid also in finite size samples. For this purpose, let us consider the response of the finite size network to the application of an external stimulus, in analogy with the results for the HMF approach shown in Sec.~\ref{sec5}. Let us synchronize a fraction equal to $S = 0.3$ of the excitatory nodes, following the same procedure of the HMF simulations.
Now, the stimulus perturbation can be measured through the difference between the excitatory fields produced by the non-synchronized and synchronized nodes respectively, as follows: 
\begin{equation}
	\label{form:stim_signal}
	\Delta Y_{EE}(t) = S ~Y_{EE}^{not-st}(t) - (1-S) ~Y_{EE}^{st}(t), 
\end{equation}
where $Y_{EE}^{not-st}(t)$ and $Y_{EE}^{st}(t)$ are computed according to the first sum in Eq.~(\ref{Y_def_FS}), limiting it to the excitatory neurons which were respectively not stimulated and stimulated. We also add the normalization factors $S$ and $1-S$, as the first and the second field are produced by a different number of neurons: in this way, before the stimulus, $\Delta Y_{EE}(t)$ fluctuates around zeros. The temporal evolution of such variable is plotted with orange and red lines on the raster plots in Figs.~\ref{fig:stim_fs}(a)--\ref{fig:stim_fs}(d). As we have periodic signals, in order to better describe their decays, we compute the average value for each period and we plot the results in the upper panel of Fig.~\ref{fig:stim_fs}.

In the initial dynamical regime of partial synchronization ($f_I = 0.1$), the perturbation produced by the stimulus decays fast and after few oscillations the system returns to the initial configuration [see the raster plots in Figs.~\ref{fig:stim_fs}(a) and \ref{fig:stim_fs}(b)]. In the metastable and in the totally synchronous regime, when the inhibitory fraction is around the balance value ($f_I = 0.214$ and $f_I = 0.23$), the time required to return to the original synchronization level is longer and the signal decay is slower. Then the decay rate increases again if we consider the asynchronous states after the balance regime ($f_I = 0.28$ for example). 
These results point out that the long response time of the system to an external stimulus is an effect which is present not only in the HMF approach but also in finite networks, though in this case the balance regime is asymptotically asynchronous.

\section{\small CONCLUSION AND PERSPECTIVES}
\label{sec7}
We have studied the role of inhibitory hubs for the synchronization and input processing of a neural network with short term plasticity. We have used a simple correlation rule for the network generation, in which input and output connectivities are the same for each neuron: this allows us to emphasize the control role of highly connected neurons, both in input and in output direction. We apply a heterogeneous mean-field approach to the finite size network dynamics, that lets us speed up numerical computations and highlight the role of neuronal connections distributions. 

In this modeling approach, hub inhibitory neurons turn out to be strongly effective to drive the network synchronization, even in the case in which their relative number is small with respect to the excitatory component. Indeed, if their hub character (their connectivity with respect to that of the excitatory population) is high, a small fraction of inhibitory neurons ($f_I\sim 0.1$) is able to increase network synchronization, leading the majority of neural population (actually the excitatory component) to synchronize at a certain frequency, which results to be equal to that of the isolated LIF neuron. Around this regime of complete synchronization we find an interesting metastable dynamical phase.
In this dynamical regime, excitatory neurons spike in clusters of different size, the composition of each cluster changes in time and it is dependent on initial conditions. Interestingly, this dynamical regime is the most appropriate for storing information of an external input received by the network. The hub nature of inhibitory neurons allows us to control the overall population dynamics even when the network is mainly composed by excitatory neurons. In fact, in the case of the absence of inhibitory hubs, i.e., when excitatory and inhibitory neurons have the same average connectivity, the totally synchronous regime appears only at high fraction of inhibitory nodes. On the other hand, as soon as the connectivity of the inhibitory component grows with respect to the excitatory one (i.e., the inhibitory neurons hub character increases), the metastable dynamical phase appears for lower and lower inhibitory fractions. 
Eventually, for very high values of $f_I$, the dynamics of the neural population is completely asynchronous. As a result, the fraction of inhibitory neurons and their hub character are two crucial ingredients for observing a wide range of dynamical phases and drive the overall network synchronization. 
In our approach, the coincidence between input and output connectivities for each neuron can have strong effects on the overall dynamics. In future steps, we plan to investigate the case in which such a correlation is reduced and hub neurons are identified only by a high input or output connectivity. 

%

\end{document}